\documentclass[aps,twocolumn,showpacs,prl]{revtex4}

\usepackage{hyperref}
\usepackage{amstext}
\usepackage{amssymb}
\usepackage{esint}

\begin{document}

\title{Maxwell-Chern-Simons Hydrodynamics for the Chiral Magnetic Effect}

\author{\c{S}ener \"{O}z\"{o}nder}

\email{ozonder@physics.umn.edu}

\affiliation{School of Physics and Astronomy, University of Minnesota, Minneapolis,
Minnesota 55455, USA}
\begin{abstract}
The rate of vacuum changing topological solutions of the gluon field,
sphalerons, is estimated to be large at the typical temperatures of
heavy-ion collisions, particularly at the Relativistic Heavy Ion Collider.
Such windings in the gluon field are expected to produce parity-odd
bubbles, which cause separation of positively and negatively charged
quarks along the axis of the external magnetic field. This chiral
magnetic effect can be mimicked by Chern-Simons modified electromagnetism.
Here we present a model of relativistic hydrodynamics including the
effects of axial anomalies via the Chern-Simons term. 
\end{abstract}



\maketitle

\section*{Introduction}

The ground state of strong interactions is believed not to break
$\mathcal{CP}$ invariance owing to a theorem by Vafa and Witten \cite{Vafa:1984xg}.
However, it has been suggested that the Vafa-Witten theorem does not
preclude formation of metastable $\mathcal{P}$- and $\mathcal{CP}$-odd
domains \cite{Kharzeev:1998kz,Cohen:2001hf,Kharzeev:2001ev,Kharzeev:2000na}.
These $\mathcal{P}$-odd bubbles are expected to occur randomly anywhere
in space-time in hot quark-gluon plasma \cite{Warringa:2009rw}, and
they have the property that $\langle\vec{E}^{a}\cdot\vec{B}^{a}\rangle\neq0$,
where the fields are color electric and magnetic fields \cite{Voloshin:2000xf}. 

$\mathcal{P}$-odd bubbles may yield observable effects on quarks
in the presence of a strong external (electromagnetic) magnetic field
because of the winding of the gluon field. The necessary setup for
observation might be achieved in heavy-ion collisions when two nuclei
collide with a nonzero impact parameter, which results in a strong
magnetic field normal to the collision plane. 

In the deconfinement phase of quark-gluon plasma, the chiral limit
($m_{\text{quark}}=0$) is a reasonable approximation. In this limit
quarks have definite helicity eigenstates as the left-handed and right-handed
helicity states do not mix in the absence of a mass term in the Lagrangian.
Accordingly, the momentum and spin will be parallel for the right-handed
massless quarks and antiparallel for the left-handed ones. In the
meantime, if the gauge field has a nonvanishing winding number, the
chirality will not be conserved owing to the axial anomaly, and vacuum-changing topological solutions will create net chirality. If there
are more left-handed or right-handed quarks than the other, then when
they align their spins parallel or antiparallel to the external magnetic
field according to their charge, charge separation will occur in the
direction of the magnetic field. As a result, a net electric current
perpendicular to the collision plane will be produced. This is called
the chiral magnetic effect (CME)  \cite{Kharzeev:2004ey,Kharzeev:2007tn,Kharzeev:2007jp,Fukushima:2008xe,Voloshin:2010ju}. Charge separation creates an electric
dipole moment and it causes $\mathcal{CP}$ violation. 

As chirality nonconservation is purely a nonperturbative effect, any
evidence for this effect in the data would be a valuable observation
of the nontrivial topology of Yang-Mills theories. The CME has been
verified in lattice calculations \cite{Buividovich:2009wi,Buividovich:2009ih,Buividovich:2008wf}.
According to the STAR Collaboration, data from RHIC indicate that
charge separation has indeed been observed \cite{Voloshin:2008jx,Caines:2009yu,Voloshin:2009hr,Selyuzhenkov:2005xa}.

\section*{Vacuum of Yang-Mills Theories}

The vacuum of non-Abelian gauge theories are infinitely degenerate;
there are infinitely many pure gauge configurations of the gauge field.
Different vacua are distinguished by an integer winding number, defined
as

\begin{equation}
Q_{\text{w}}=\frac{g^{2}}{32\pi^{2}}\int d^{4}xG^{a\mu\nu}\widetilde{G}{}_{\mu\nu}^{a},\end{equation}
where $\widetilde{G}_{\mu\nu}^{a}=\frac{1}{2}\epsilon^{\mu\nu\alpha\beta}G_{\alpha\beta}^{a}$
is the dual field strength.

No small perturbation around the minimum of a vacuum can change the
winding number; perturbation theory is blind to the many-vacua structure
of the non-Abelian theory. Instantons are finite solutions of the Euclidian
action and they are localized in space-time with the feature of tunneling
between two different vacua. Their effect, however, is suppressed
exponentially at any temperature. Another possibility of the vacuum
changing solution is the sphaleron. It interpolates from one vacuum
to another by hopping over the energy barrier, whose height is on
the order of $\Lambda_{QCD}$. This is possible at high temperatures
because there will be enough energy to surmount the barrier without
tunneling. Although sphalerons originally appeared in the electroweak
theory, their importance also arises in quark-gluon plasma \cite{McLerran:1990de}.
The approximate sphaleron rate is (see Ref. \cite{Warringa:2008kv}, and references therein) 

\begin{equation}
\Gamma=\frac{d N}{d^{3}x dt}\thickapprox386\alpha_{s}^{5}T^{4},\label{eq:sphaleron rate}\end{equation}
where $\alpha_{s}=g^{2}/4\pi$ and $T$ is the temperature. As sphalerons
do not tunnel, the rate is not suppressed at high temperatures and
an appreciable number of transitions is expected.

\section*{QED Coupled to QCD with Topological Charge}

The charge separation may be examined from a more formal viewpoint.
The Lagrangian for QED coupled to QCD with topological charge is given
by \cite{Kharzeev:2009fn}

\begin{eqnarray}
{\cal L}_{{\rm QCD+QED}}=\,\,\,\,\,\,\,\,\,\,\,\,\,\,\,\,\,\,\,\,\,\,\,\,\,\,\,\,\,\,\,\,\,\,\,\,\,\,\,\,\,\,\,\,\,\,\,\,\,\,\,\,\,\,\,\,\,\,\,\,\,\,\,\,\,\,\,\,\,\,\,\,\,\,\,\,\,\,\,\,\,\,\,\,\,\,\,\,\,\,\,\,\nonumber \\
\sum_{f}\bar{\psi}_{f}\left[i\gamma^{\mu}(\partial_{\mu}-igA_{\mu}^{a}t^{a}-iq_{f}A_{\mu})-m_{f}\right]\psi_{f}\nonumber \\
-\frac{1}{4}G^{a\mu\nu}G_{\mu\nu}^{a}-\frac{\theta}{32\pi^{2}}g^{2}G^{a\mu\nu}\widetilde{G}{}_{\mu\nu}^{a}-\frac{1}{4}F^{\mu\nu}F_{\mu\nu},\label{eq:qcdqed}\end{eqnarray}
\begin{eqnarray*}
\end{eqnarray*}
where $A_{\mu}^{a}$ is the gluon field, $A_{\mu}$ is the electromagnetic
field, and $G_{\mu\nu}^{a}$ and $F_{\mu\nu}$ are the gluon and electromagnetic
field strength tensors, respectively. Although the term $F^{\mu\nu}\widetilde{F}_{\mu\nu}$
does not exist in the Lagrangian, it will be induced by the topological
charge of the gluon field via quark loops \cite{Asakawa:2010bu}.
To focus on the electromagnetic sector of this theory we
can start with the effective Maxwell-Chern-Simons Lagrangian \cite{Sikivie:1984yz,Wilczek:1987mv,Carroll:1989vb}:

\begin{equation}
{\cal L}_{{\rm MCS}}=-\frac{1}{4}F^{\mu\nu}F_{\mu\nu}-A_{\mu}J^{\mu}-\frac{c}{4}\theta F^{\mu\nu}\widetilde{F}_{\mu\nu},\label{eq:effective}\end{equation}
where $\theta=\theta(\vec{x},t)$ will mimic ${\cal P}$-odd bubbles,
$c=\sum_{f}q_{f}^{2}e^{2}/(2\pi^{2})$ and $J^{\mu}$ is the electric
current of the quarks. The last term in Eq. (\ref{eq:effective}) is not
a total derivative and does not vanish because $\theta$ is a field
and is determined by fluctuations of the topological charge. The equation
of motion following from (\ref{eq:effective}) is

\begin{equation}
\partial_{\mu}F^{\mu\nu}=J^{\nu}-c\tilde{F}^{\sigma\nu}\partial_{\sigma}\theta.\label{eq:mcs-eom}\end{equation}
Together with $ $$\partial_{\mu}\tilde{F}^{\mu\nu}=0$, the set of
Maxwell-Chern-Simons equations can be written as 

\begin{equation}
\vec{\nabla}\times\vec{B}-\frac{\partial\vec{E}}{\partial t}=\vec{J}+c\left(\dot{\theta}\vec{B}+\vec{\nabla}\theta\times\vec{E}\right),\label{eq:max1}\end{equation}

\begin{equation}
\vec{\nabla}\cdot\vec{E}=\rho-c\vec{\nabla}\theta\cdot\vec{B},\label{eq:max2}\end{equation}

\begin{equation}
\vec{\nabla}\times\vec{E}+\frac{\partial\vec{B}}{\partial t}=0,\end{equation}

\begin{equation}
\vec{\nabla}\cdot\vec{B}=0.\end{equation}
These reduce to Maxwell's equations when the field $\theta$ is 0.
The modification of electromagnetism can be attributed to the vacuum
changing solutions in the full $\text{QED}\times\text{QCD}$ theory.

\section*{Maxwell-Chern-Simons Hydrodynamics}

A very strong magnetic field is created in heavy ion collisions when
two highly charged nuclei collide with a nonzero impact parameter.
Considering typical RHIC parameters, the magnetic field generated
normal to the collision plane is of the order $eB\thicksim(10-100\mbox{ MeV})^{2}$
during the first moments ($\tau\thicksim1$ fm/c) of the collision.
This magnetic field is even stronger than the ones in magnetized neuron
stars and magnetars, where typical magnetic fields are about $eB\thickapprox(2\mbox{ MeV})^{2}$,
or $10^{10}$ T. 

Relativistic hydrodynamics is the prime tool being used to describe
the time evolution of quark-gluon plasma. Let us start with the
conservation of energy and momentum to derive the hydrodynamic model
of quarks coupled to Chern-Simons modified electromagnetism. 

\begin{equation}
\partial_{\mu}(T^{\mu\nu}+\Theta^{\mu\nu})=0.\label{eq:fluidcons}\end{equation}
Here the energy-momentum tensor of the ideal fluid is given by

\begin{equation}
T^{\mu\nu}=(\epsilon+P)u^{\mu}u^{\nu}-g^{\mu\nu}P,\end{equation}
where $\epsilon$ and $P$ are the local energy density and pressure.
The fluid four-velocity is $u^{\mu}=(\gamma,\gamma\mathrm{\vec{v})}$,
and in the rest frame of the fluid it reduces to $u^{\mu}=(1,0,0,0)$.
We use the natural units $\hbar=c=k_{B}=1$ and the metric $g^{\mu\nu}=\mbox{diag}(+1,-1,-1,-1)$.
The energy-momentum tensor of the free electromagnetic field is given
by 

\begin{equation}
\Theta^{\mu\nu}=F_{\lambda}^{\mu}F^{\lambda\nu}+\frac{1}{4}g^{\mu\nu}F_{\alpha\beta}F^{\alpha\beta}.\end{equation}
Its divergence can be found by using the equation of motion (\ref{eq:mcs-eom}),

\begin{equation}
\partial_{\mu}\Theta^{\mu\nu}=-F^{\nu\lambda}J_{\lambda}-cF^{\nu\lambda}\widetilde{F}_{\lambda\sigma}\partial^{\sigma}\theta.\end{equation}
Substituting this into Eq. (\ref{eq:fluidcons}) gives

\begin{equation}
\partial_{\mu}T^{\mu\nu}=F^{\nu\lambda}J_{\lambda}+cF^{\nu\lambda}\widetilde{F}_{\lambda\sigma}\partial^{\sigma}\theta.\label{eq:mcs-hydro}\end{equation}

The electric current of the quarks in the fluid is defined as $J^{\mu}=nu^{\mu}$,
where $n$ is the net electric charge density. It is conserved, as
can be seen by taking the divergence of both sides of (\ref{eq:mcs-eom})

\begin{equation}
\partial_{\mu}J^{\mu}=0.\end{equation}

By using the definitions $E^{\mu}\equiv F^{\mu\nu}u_{\nu}$ and $B^{\mu}\equiv\widetilde{F}^{\mu\nu}u_{\nu}$,
Eq. (\ref{eq:mcs-hydro}) can be written as

\begin{equation}
\partial_{\mu}T^{\mu\nu}=nE^{\nu}-cE^{\lambda}B_{\lambda} \partial^{\nu}\theta ,\label{eq:mcs-hydro-EB}\end{equation}
where $E^{\lambda}B_{\lambda}=-\vec{E}\cdot\vec{B}.$ We refer to
the right-hand side of Eq. (\ref{eq:mcs-hydro-EB}) as the Lorentz-Chern-Simons
force density. 

Now we calculate the entropy production in the system by contracting
Eq. (\ref{eq:mcs-hydro-EB}) with $u_{\nu}$. Using the fact that $u_{\nu}E^{\nu}=0$,
current conservation, and the normalization condition of the four-velocity
$u^{\mu}u_{\mu}=1$, we find

\begin{equation}
nu^{\mu}\left(\partial_{\mu}\left(\frac{\epsilon+P}{n}\right)-\frac{1}{n}\partial_{\mu}P\right)=-cE^{\lambda}B_{\lambda}u_{\nu}\partial^{\nu}\theta.\label{eq:entro1}\end{equation}
By using the Gibbs relation,

\begin{equation}
d \left( \frac{\epsilon+P}{n} \right)=Td \left( \frac{\sigma}{n} \right)+\frac{1}{n}dP,
\end{equation}
Eq. (\ref{eq:entro1}) can be transformed into

\begin{equation}
\partial_{\mu}s^{\mu}=-\frac{1}{T}cE^{\lambda}B_{\lambda}u_{\nu}\partial^{\nu}\theta,\label{eq:entro2}\end{equation}
where $\sigma$ is the entropy density, and the entropy current is
$s^{\mu}=\sigma u^{\mu}$. Here we see that the Chern-Simons term
can increase or decrease the entropy as it can be positive or negative.
Although it may appear to be a violation of the second law of thermodynamics,
we point out that entropy can decrease locally. However, the change
in the entropy of the whole system cannot be negative. This situation
is similar to that of Maxwell's demon, where the demon can collect
fast-moving and slow-moving gas molecules on two separate sides of
the container by controlling the door in the middle of the container
and, consequently, decrease the entropy of the gas by driving the system
into more ordered phases. However, this does not violate the second
law of thermodynamics. The entropy produced by the demon cannot be
less than the entropy decrease of the gas, which means that the total
entropy change of the whole system, the gas and the demon, cannot
be less than 0. In Eq. (\ref{eq:entro2}) we isolated the external
field $\theta(\vec{x},t)$ and calculated the entropy for the rest
of the system. 

Discussion of the entropy also relates to the vacuum structure
of the $SU(N)$ gauge theories. It is well known that the vacuum of
non-Abelian gauge theories is infinitely degenerate. Topological solutions,
such as instantons and sphalerons, interpolate between different vacua,
and what enumerates that is the winding number. As positive and
negative change in the winding number are equally possible, its behavior
is completely determined by a one-dimensional random walk. Therefore,
a net change in the winding number can happen in a particular event
by means of fluctuations. The parallel electric and magnetic fields
produced by the term $ $$G^{a\mu\nu}\widetilde{G}{}_{\mu\nu}^{a}\sim\vec{E}^{a}\cdot\vec{B}^{a}$
correspond to increase or decrease in the entropy locally; however,
when infinitely many degenerate vacua are considered, the total entropy
does not decrease. We note that in Ref. \cite{Son:2009tf} an extra term
was added to the current to cure the negative entropy change problem.
Here we take a different approach. 

To calculate the entropy of the whole system, we use
the modified Gibbs relation (see also Ref. \cite{Aguiar:2003pp}):

\begin{equation}
d\left(\frac{\epsilon+P}{n}\right)=Td\left(\frac{\sigma}{n}\right)+\frac{1}{n}dP+\frac{1}{n}R_{\theta}d\theta,\label{eq:mod-gibbs}\end{equation}
where the generalized force $R_{\theta}$ is defined as 

\[
R_{\theta}\equiv\left(\frac{\partial\Omega}{\partial\theta}\right)_{T,\mu},\]
and $\Omega$ is the thermodynamic potential density. In terms of
the partition function of the model given in Eq. (\ref{eq:effective}) ,

\begin{equation}
Z=\int[{\cal D}q][{\cal D}\bar{q}][{\cal D}A_{\mu}]\text{exp}\left(\int_{0}^{\beta} d \tau\int d^{3}x{\cal L}\right)\end{equation}
where $\beta=1/T$, the thermodynamic potential can be written as

\begin{equation}
\Omega=-P=\frac{T}{V}\mbox{ln}Z.\end{equation}
The term $-\frac{c}{4}\theta F^{\mu\nu}\widetilde{F}_{\mu\nu}=c\theta\vec{E}\cdot\vec{B}$
in the Lagrangian, Eq. (\ref{eq:effective}), involves fields such as $\vec{B}$,
which is generated by the colliding nuclei, and $\vec{E}$, which is
induced by the field $\theta(\vec{x},t).$ Here we assume that the
fields vary so slowly in space and time that they can be taken outside
the integration to give 

\begin{eqnarray}
Z & = & \text{exp}\left(-\frac{V}{T}c\theta E^{\lambda}B_{\lambda}\right)\nonumber \\
 &  & \times\int[{\cal D}q][{\cal D}\bar{q}][{\cal D}A^{\prime}_{\mu}]\text{exp}\left(\int_{0}^{\beta} d \tau\int d^{3}x{\cal L}^{\prime}\right)\end{eqnarray}
This assumption is no better or worse than the application of hydrodynamics
to heavy-ion collisions, where it is necessary to assume that the matter
is locally thermalized and that the temperature and flow velocity
vary slowly in space and time. The functional integration in $Z$
involves fluctuations $A^{\prime}_\mu$ around the externally imposed $\vec{B}$ field
and the induced $\vec{E}$ field. Hence $R_{\theta}=-cE^{\lambda}B_{\lambda}.$
Now we rewrite Eq. (\ref{eq:entro1}) as

\begin{equation}
nu^{\mu}\left(\partial_{\mu}\left(\frac{\epsilon+P}{n}\right)-\frac{1}{n}\partial_{\mu}P+\frac{1}{n}cE^{\lambda}B_{\lambda}\partial_{\mu}\theta\right)=0.\end{equation}
Using Eq. (\ref{eq:mod-gibbs}) we deduce that entropy is conserved:

\begin{equation}
\partial_{\mu}s^{\mu}=0.\label{eq:entropy-con}\end{equation}
In this calculation we used ideal hydrodynamics and ignored the dissipative
terms. It is well-known that the inclusion of dissipative terms in
$T^{\mu\nu}$ and $J^{\mu}$ does not result in a decrease in entropy.

In Eq. (\ref{eq:entro1}) we calculated the component of the energy-momentum
conservation equation in the direction of $u_{\nu}$. Now we
find the projection of Eq. (\ref{eq:mcs-hydro-EB}) perpendicular to $u_{\nu}$.
Multiplying both sides of Eq. (\ref{eq:mcs-hydro-EB}) by the projection
tensor $\Delta_{\nu}^{\alpha}=\delta_{\nu}^{\alpha}-u^{\alpha}u_{\nu}$,
we get

\begin{eqnarray}
(\epsilon+P)u^{\mu}\partial_{\mu}u^{\alpha}-(\delta_{\nu}^{\alpha}-u^{\alpha}u_{\nu})\partial^{\nu}P=nE^{\alpha} \nonumber \\
-cE^{\lambda}B_{\lambda} (\delta_{\nu}^{\alpha}-u^{\alpha}u_{\nu})  \partial^{\nu}\theta.\label{eq:mcs-euler}\end{eqnarray}

Defining the comoving
derivative as $D=u_{\mu}\partial^{\mu}$, the normal derivative as $D^\alpha=\Delta_{\nu}^{\alpha} \partial^{\nu}$, and using the definition $w=\epsilon+P$, we can write Eqs. (\ref{eq:entro1}) and (\ref{eq:mcs-euler})
as

\begin{eqnarray}
D\epsilon+w\partial_{\mu}u^{\mu} & = & -cE^{\lambda}B_{\lambda}D\theta,\\
wDu^{\alpha}-D^\alpha P & = & nE^{\alpha}-cE^{\lambda}B_{\lambda} D^{\alpha}\theta.\label{eq:euler}\end{eqnarray}
Equation (\ref{eq:euler})  is reminiscent of the nonrelativistic
Euler's equation.

\section*{Conclusion}

We have considered Chern-Simons modified electromagnetism in the context
of the Chiral Magnetic Effect. We formulated the hydrodynamic model
for the purpose of incorporating axial anomalies into relativistic
hydrodynamics. The equations we found should be solved numerically
by using realistic values for the magnitude of the external magnetic
produced in heavy ion collisions and the size and rate of the sphaleron
transitions. This work is now in progress. 
\begin{acknowledgments}
We thank J. I. Kapusta for discussions and very useful comments on
the manuscript. We also thank D. T. Son for clarifying certain points
in his paper. This work was supported by the US Department of Energy
(DOE) under grant DE-FG02-87ER40328.
\end{acknowledgments}

\end{document}